\documentstyle[12pt]{article}
\newcommand{\bq}{\begin{equation}}
\newcommand{\eq}{\end{equation}}
\newcommand{\ra}{\rightarrow}
\newcommand{\ov}{\overline}
\begin{document}

\begin{flushright}
{\bf BINP 98-62}\\
hep-th/9808093\\
\end{flushright}
\vspace{.5cm}

\begin{center}{\bf PROPERTIES OF ${\cal N}=1$ SUSY YANG-MILLS VACUUMS\\ 
AND DOMAIN WALLS}\\
\vspace{.5cm}
{\bf Victor Chernyak}\footnote{e-mail: chernyak@inp.nsk.su}\\
\vspace{.5cm}
Budker Institute of Nuclear Physics, 630090 Novosibirsk-90, Russia
\end{center}

\vspace{1.cm}

\begin{abstract}
It is shown that there is no chirally symmetric vacuum state in the
${\cal N}=1$\, supersymmetric Yang-Mills theory. 
The values of the gluino condensate and the vacuum energy density are
found out through a direct instanton calculation. A qualitative picture of
domain wall properties is presented, and a new explanation of the
phenomenon of strings ending on the wall is proposed. 
\end{abstract}

\hspace{.5cm} {\bf 1.} \hspace{.5cm} The ${\cal N}=1$ supersymmetric 
Yang-Mills theory (SYM) partition function is ( Q is the topological charge):
\bq
Z=\sum_k \int dA_\mu d\lambda d{\bar \lambda}\,\delta(Q-k)\,\exp \left 
\{\frac {i}{4g_o^2}\int dx d\theta \,\frac{1}{2}\,W_\alpha ^2 +h.c.
\right \}\,.
\eq
Let us integrate it now over the gluon and gluino fields, but with the chiral
superfield $W^2$ being fixed. Proceeding as in \cite{I}, 
one obtains the partition
function for the chiral superfield $\Omega=(W^2/32\pi^2 N_c)$ in the form:
\footnote{
 The notation $\oint$ means that the quantum loop contributions 
of the superfield 
$W^2$ have also been integrated out, so that the exact correlators of the
superfield $W^2$ are obtained from eq.(2) using the tree diagrams only. See
\cite{I} for more detail.
}
$$
Z=\sum_{k}\oint d\Omega\,d{\bar \Omega}\,\delta (Q-k)\,\exp \{i\int dx L\},$$
\bq
\frac{1}{N_c^2}L=\left \{\frac{1}{2}\int d\theta\, \Omega\ln\left (\frac
{\Omega}{e\Lambda^3}\right )+h.c.\right \}+\int d\theta d{\bar \theta}\, 
M(\Omega,{\overline \Omega}, D^n\Omega, \overline {D^n\Omega}, \dots).
\eq  
(where $D$ and $\ov D$ are the superderivatives). The F-term of the 
Lagrangean in eq.(2), which accounts for all (super) anomalies, coincides 
with the F-term of the well known Veneziano-Yankielowitz (VY) \cite{VY} 
effective Lagrangean. The difference is that 
the meaning of the of the word "effective" was not quite clear for the 
VY-Lagrangean, as well as its connection with the original fundamental 
Lagrangean. In our approach (see \cite{I} for detail) its 
connection with the fundamental YM-Lagrangean and its meaning become clear: 
it is the exact generating functional of the (one particle irreducible)
Green functions of the field $\Omega$. 

The D-term in eq.(2) is 
nonanomalous and depends both on the field $\Omega$ and its superderivatives. 
(For our purposes, we will ignore in what follows all fermionic components 
of $\Omega$ and all terms with usual space-time derivatives). 

We will show in this section that the chirally symmetric vacuum state 
obtained by A. Kovner and M. Shifman \cite{KS} (KS-vacuum with $\langle 0|
\lambda\lambda|0\rangle=0$) is an artefact of using the total VY-Lagrangean,
i.e. with the D-term in eq.(2) chosen in the simplest form:
\bq
M=const\,\left ({\bar \Omega}\Omega\right )^{1/3}\,.
\eq
In what follows, we prefer to deal with the usual component fields:
\bq \Omega=(\sigma,\,\theta^2\chi),\quad \sigma=\frac{\lambda\lambda}{32
\pi^2 N_c}\,,\quad \chi=S+iP=\frac{GG+iG\tilde G}{32\pi^2 N_c}\,\,,
\eq
so that the VY-potential takes the form:
\bq
\frac{1}{N_c^2}U=\frac{1}{2}\left \{\left (S+iP\right )\ln \left (\frac
{\sigma}{\Lambda^3}\right )+h.c.\right \}-C_o\frac{S^2+P^2}{|\sigma|^{4/3}}\,.
\eq
With this form, there are $N_c$ chirally asymmetric vacuum states:
\bq
{\bar \sigma}_n={\langle 0|\sigma|0\rangle}_n {\sim \langle 0 |\lambda\lambda|
0 \rangle}_n \sim \Lambda^3 \exp \left \{i\frac{2\pi n}{N_c} \right \},\,
n=0,...,N_c-1,
\eq
corresponding to the spontaneously broken residual axial symmetry and besides, 
as emphasized by A. Kovner and M. Shifman \cite{KS}, there is also the 
chirally symmetric vacuum solution: 
\bq
{\langle 0|\lambda\lambda|0\rangle}_o =0.
\eq
Let us point out first that two solutions, eq.(6) and eq.(7), are not on
equal footing. Because we know (from the Witten index) that SUSY is unbroken,
we are ensured that ${\ov S}=
\langle 0|S|0\rangle \ra 0$. So, if $|{\ov \sigma}|\neq 0$, it is
sufficient to use eq.(5) to find out the value of ${\ov \sigma}$, as higher
order terms, like $S(S/|\sigma|^{4/3})^k$, are of no importance in this case.
If $|{\ov \sigma}|\ra 0$ however, all higher order terms become of importance
and we can not believe, in general, the results obtained from eq.(5). If the
VY-potential were exact, the KS-solution will survive. But really, the term
$S^2/|\sigma|^{4/3}$ in eq.(5) is only the first term in the expansion in 
powers of $(S/|\sigma|^{4/3})^k$. So, the KS-solution is not selfconsistent 
in this respect and we need to know, in particular, the behaviour of the 
potential at $z=S/|\sigma|^{4/3}\ra \infty$. 

To find it out, let us write first a general form of the potential in eq.(2)
($\sigma=\rho \exp \{ i \phi \})$:
\bq
\frac{1}{N_c^2}U(\sigma,\chi)=S\ln\left (\frac{\rho}{\Lambda^3} \right )-
(\phi-\frac{\theta}{N_c})P+Sf_{1}\left (\frac{S}{\rho^{4/3}}\,,\,
\frac{P}{S}\right ).
\eq
Let us add now to eq.(8) the gluino mass term:
\bq
\frac{1}{N_c^2}\Delta U=-m_o\,\rho\cos(\phi)\,\,,
\eq
where $m_o$ is the renormalization group invariant mass parameter.
\footnote{
Changing the phase of $m_o$ is equivalent to a redefinition of $\theta$ in
eq.(8). So, it is convenient to choose $m_o$ in eq.(9) to be real and
positive.
}
This addition of $\Delta U$ is legitimate as our Lagrangean was obtained
integrating out all degrees of freedom, but with all components of the
$\Omega$-superfield fixed, and because 
the gluino mass, $m_o$, can be considered 
as a source for the field $\lambda\lambda$ (see \cite{I} for more detail). 
Now, at large $m_o\ra \infty$, the heavy gluino will decouple leaving us
with the pure YM theory and we know how it decouples, from the 
renormalization group. In this region (see below): ${\ov S}=O(m_o^{8/11}),
\, {\ov \rho}=O(m_o^{-3/11})$, so that ${\ov S}/{\ov \rho}^{4/3}=O(m_o^{12
/11})\ra \infty$. Therefore, this will allow us to find out the asymptotic
behaviour of $f_1$ in eq.(8).

As the gluino contribution to $b_o=3=(11/3\,-\,2/3)$ is $(-2/3)$, it is not
difficult to check that the function $f_1$ in eq.(8) has to have the
asymptotic behaviour:
\bq
f_{1}\left (\frac{S}{\rho^{4/3}},\,\frac{P}{S}\right )\ra \frac{1}{4}\ln 
\left (\frac{S}{\rho^{4/3}} \right )+ f_{2}\left (\frac{P}{S} \right ),\quad
\frac{S}{\rho^{4/3}}\ra \infty.
\eq
In this case, integrating out the $\rho$ and $\phi$ fields, one has:
\bq
m_o{\ov \rho}\,e^{i{\ov \phi}}=\left (\frac{2}{3}S+iP \right ),
\eq
and $U(S,P)$ takes the form:
\bq
\frac{1}{N_c^2}U(S,P)=\frac{11}{12}S\ln \left (\frac{S}{\Lambda^4_{YM}}
\right )+\frac{\theta}{N_c}P+S f{_3}\left (\frac{P}{S}\right );\,\, 
\Lambda_{YM}= \Lambda^{9/11}\,m_o^{2/11},
\eq
as it should be.
\footnote{
Another way to check eq.(10) is to recall that eq.(11) can be obtained 
through a direct calculation of the heavy gluino loop in the gluon 
background, and is directly related to the trace and axial anomalies.}

Now, we are ready to check the existence of the KS-solution: 
${\bar \rho}\ra 0$.
We can distinguish three cases (we take $\theta=0,\,{\bar \phi}={\bar P}=0$,
as they are of no importance for us here).\\
a) Let ${\bar z}=({\bar S}/{\bar \rho}^{4/3})\ra 0$, so that $f_{1}(z=S/\rho^
{4/3})\sim z.$ As was pointed out above, this variant is selfcontradictory
at $\bar \rho \ra 0$, as $\partial U/\partial S=0$ leads to: ${\bar z}
\sim \ln ({\Lambda^3}/{\bar \rho})\ra \infty.$\\
b) Let ${\bar z}\ra z_o=const \neq 0.$ Then (barring 
pathological singularities) the
saddle point equations are: $z_o f_{1}^{\prime}(z_o)=3/4;\,\, \ln(\Lambda^3
/{\bar \rho})=f_{1}(z_o)+z_o f_{1}^{\prime}(z_o).$ The first equation shows 
that $f_{1}^{\prime}(z)$ (and so $f_{1}(z)$) is nonsingular at $z=z_o$, but 
we are in trouble then with the second equation at ${\bar \rho}\ra 0$.\\
c) Finally, let ${\bar z}\ra \infty$, so that $f_{1}(z)\ra (1/4)\ln z$. This 
case is also in trouble, as $\partial U/\partial S=0$ leads to $({\bar S}/
{\bar \rho})=O(1/{\bar \rho}^{11/3})\ra \infty$ at ${\bar \rho}\ra 0$, while
$\partial U/{\partial \rho}=0$ leads to $({\bar S}/{\bar \rho})\ra 0$.

On the whole, we conclude that there is no chirally symmetric vacuum state
in ${\cal N}=1$ SYM, so that the residual axial symmetry is spontaneously
broken in all vacuum states.

\hspace{.5cm}{\bf 2.}\hspace{.5cm}
We will show now that a spontaneous breaking of the residual axial
symmetry and the value of the gluino condensate can be obtained in a quite
different way, through a direct calculation of the instanton contributions 
into the partition function. With this purpose, let us return to the original
partition function, eq.(1), add the gluino mass term with a small but finite
mass $m_o$ to the action, and consider the instanton contributions.

It has been shown in \cite{ins} that, under a special choice of the 
collective coorditates, the n-instanton
contribution splits up into $n N_c$ "instantonic
quarks". In our case of ${\cal N}=1\,\, SYM$, the result is especially simple.
Because all nonzero mode contributions cancel exactly between the gluon and 
gluino contributions, there remains no residual interaction between these
instantonic quarks. For instance, the n=1 instanton contribution takes the
form ( $b_o=3,\, m_{\theta}=m_o\exp\,(i\,\theta/N_c)\,)$:
\bq
Z_1=\int dx_1 ... dx_{N_c}\,\frac{1}{N_c!}\left [\,N_c^2\,\frac{m_{\theta}}{2}
\,\Lambda^{b_o}\,\right ]^{N_c}\,,
\eq
where: $x_i$ are the collective coordinates (the  
positions of the instantonic quarks), and $m_{\theta}$ 
is due to the gluino zero modes. The n-instanton contribution
is exactly of the same form, so that summing up over n 
(and adding antiinstantons) one obtains the partition function in the form:
$$
Z_{tot}=Z\,Z^*\,, \quad Z=\frac{1}{N_c}\sum_{k=0}^{N_c-1}e^{I(k)}\,,
$$
\bq
I(k)=\int dx\, N_c^2\left\{\frac{m_{\theta}}{2}\,\Lambda^3 \left [1+
O(|m_{\theta}|^2)\right ]\exp\left (i\,\frac{2\pi k}{N_c} \right )+
O(|m_{\theta}|^2)\right\}.
\eq
Here, the factor $Z_{N_c}(k)=\exp\left\{i\,2\pi k/N_c \right \}$
appeared because we have extracted the $N_c$-th power root from unity, 
when going from eq.(13) to eq.(14). It plays the role of the "neutralizator",
i.e. when $\exp\{I(k)\}$ in eq.(14) is expanded back into a power series, it
ensures that instantonic quarks appear in the $N_c$\,-fold clusters
only (i.e. in the form of instantons). Besides, it ensures the periodicity:
$Z(\theta)=Z(\theta+2\pi l)$, which was explicit before summation over n.

We would like to emphasize that the above expression for
the action in eq. (14) is exact, within the indicated accuracy. Indeed:\\ 
a) The perturbation theory (i.e. the $Q=0$\, sector of the partition 
function) contribution is exactly zero at $m_o=0$ due to SUSY, and the
corrections from this sector start with $O(|m_{\theta}|^2 )$. This is because 
the replacement $m_o\ra -m_o$ is equivalent to changing the $\theta$\,-
angle, and the $Q=0$\,-sector is $\theta$\,-independent.\\
b) Because the one-loop $Z_Q$\,-contribution contains already the factor
$(m_{\theta})^{Q N_c}$, all higher loop corrections to it can be calculated 
with $m_o=0$, and they all cancel due to SUSY. For the same chirality 
reasons, the (relative) corrections 
in this sector also start with $O(|m_{\theta}|^2)$, 
including those which originate from disturbing the exact
cancelation between the gluon and gluino nonzero modes at $m_o=0$.\\
c) As for the instanton-antiinstanton interaction contributions, they
should not be considered as independent ones, but rather as belonging to 
perturbation theory (its asymptotic tail), in the sector with fixed 
 Q. So, they are also zero in the sense that they are accounted for 
already in the points "a" and "b" above.

Eq.(14) shows clearly that the residual axial symmetry is broken 
spontaneously in our system (in the infinit volume limit).
Indeed, before summation over n each n-instanton contribution was
invariant by itself under $\theta\ra \theta+2\pi l$, as a result of the
residual axial symmetry. But after summation, the instantonic
quarks have released and the above symmetry acts nontrivially now, 
interchanging $N_c$ branches between themselves.
As a result, because the small perturbation ($m_o\neq 0$) was introduced, 
one definite branch dominates the whole partition 
function,\, - those one which minimizes the energy (at given  
$\theta$). So, we obtain for the vacuum energy density:
\bq
E_{vac}=- N_c^2\,\Lambda^3\,\frac{1}{2}\left [\, m_{\theta}+{\overline m}_
{\theta}\,\right ]_{2\pi}+O(|m_{\theta}|^2)\,,
\eq
where the notation $[f(\theta)]_{2\pi}$ means that this function is
$f(\theta)$ at $-\pi\leq\theta\leq \pi$, and is glued then to be periodic in
$\theta\ra\theta+2k\pi$, i.e.: $[f(\theta)]_{2\pi}=\min_k\, f(\theta+2\pi k)$. 

Further, because the $O(m_o)$ term appeared in the energy, 
this shows that the order parameter (the gluino condensate) is nonzero.
Indeed, let us consider
\footnote{
We can keep $m_o$ infinitesimal but finite, and $V\ra \infty$, and to
separate out one term from the sum over "k" in eq. (16).} :
$$
(N_c\,V)^{-1}\exp \left\{\frac{i\theta}{N_c}\right \}\left [\frac{\partial 
\ln Z}{\partial (m_\theta/2)}\right ]_{m_o=0}\equiv N_c\, 
{\overline {\langle \lambda\lambda \rangle}}=
$$
\bq
=\sum_{k=0}^{N_c-1}\langle \theta, k|
\lambda\lambda|\theta, k \rangle =
\sum_{k=0}^{N_c-1}N_c\,\Lambda^3\exp \left 
\{ i\,\frac {\theta+2\pi k}{N_c}\right \}\,.
\eq
Thus, it is clear from eqs. (14),\,(16) that (at $m_o\ra0$), there are $N_c$ 
degenerate vacuum states differing by the phase of the gluino condensate:
$\langle \theta, k|\lambda\lambda|\theta, k \rangle =N_c\,\Lambda^3 \exp 
\,\{ i\,(\theta+2\pi k)/N_c\}.$

Moreover, it is possible to replace $m_o$ by the local function
$m_o(x)$ in eqs. (13) and (14). Indeed, in eq. (13) the gluino zero
mode contributions will take the form:
$$
I_o=\int dx_1\dots dx_{N_c} \int dy_1\, m_{\theta}(y_1)\dots dy_{N_c}
\,m_{\theta}(y_{N_c})\,\Pi\,, 
$$
$$
\Pi=|\psi_o^{(1)}(y_1-x_k)|^2\dots |\psi_o^{(N_c)}(y_{N_c}-x_k)|^2,
$$
where $\psi_o^{(i)}(y_i-x_k)$ means $\psi_o^{(i)}(y_i-x_1,\,y_i-x_2,\dots , 
y_i-x_{N_c})$, and $\int dy|\psi_o^{(i)}(y-x_k)|^2=1$. It is not
difficult to see that: $I_1=\int dx_1\dots dx_{N_c}\,\Pi=1$. Indeed, let us 
take temporarily $m_{\theta}=1$ and put our fields into a large 4-dimensional
Euclidean volume V, of an arbitrary form. Because $\int dy_1\dots dy_{N_c}
\,\Pi=1,\, I_o(m=1)=V^{N_c}$. Now, if $I_1$ were a nontrivial function of
the ratios like $(y_1-y_2)^2/(y_2-y_3)^2$ etc., then $I_o(m=1)$ will be of
the form: $I_o(m=1)=V^{N_c} f_{geom}$, where the function $f_{geom}$ will
depend on the geometry of our volume, and this will be a wrong answer. 
Therefore,  $I_o =\prod_{i=1}^{N_c}\int dy_i\,m_{\theta}(y_i)\,,$
and $m_{\theta}$ can be replaced by $m_{\theta}(x)$ in eqs. (14) and 
(15). While the corrections $O(|m_{\theta}|^2)$ in eq. (14) remain 
uncontrolable, it is important that there are no uncontrolable pure chiral 
corrections of the type $O(m_{\theta}^l),\, l\geq 2 $. As a result, taking
derivatives $\sim \delta/\delta\,m_{\theta}(x)$  we can
obtain even local pure chiral Green functions, like: $\langle k| \lambda
\lambda(x_1)\dots \lambda\lambda(x_l)|k \rangle$, 
and all of them will be pure constants.

There is nothing mysterious in this behaviour and it does not imply that the
theory is trivial. For instance, let us consider $\langle k|\lambda\lambda
(x)\,\lambda\lambda(0)|k \rangle,$ and let us denote: $\lambda\lambda(x)=
\exp (i\,2\pi k/N_c)\,\rho(x)\exp \,( i\phi_{k}(x))$, so that: $\langle k 
|\rho(x)|k \rangle \sim N_c\Lambda^3$ and $\langle k |\phi_{k}(x)| k \rangle
=0.$ Then $(\, \rho\exp (i \phi_k)=\sigma_k+i \pi_k \,)$:
$$ 
\langle k |\lambda\lambda(x)\,\lambda\lambda(0)|k\rangle =
\langle k |\lambda\lambda(x)|k \rangle \langle k|\lambda\lambda(0)|k \rangle+
$$
$$
+\exp\left \{\frac{i\,4\pi k}{N_c}\right \}\left \{\langle k|\sigma_{k}(x)
\sigma_{k}(0)|k\rangle_{con} -\langle k|\pi_{k}(x)\pi_{k}(0)|k \rangle _{con}
\right \} =\langle k|\lambda\lambda(0)|k\rangle ^2 ,
$$
as the nontrivial connected correlators cancel each other due to SUSY.

As was pointed out above, supposing only that the gluino 
condensate is really nonzero, it becomes legitimate to use the VY-Lagrangean 
to investigate the vacuum properties, i.e. to find out the gluino 
condensate, eq.(6) \cite{VY}, and the vacuum energy density, eq.(15) \cite
{MV}. In other words, it is not an approximation in this case as higher 
order terms are of no importance. In contrast, if we want to deal with 
some excitations, say domain walls, the VY-Lagrangean is insufficient.

\hspace{.5cm}{\bf 3.}\hspace{.5cm} Because there is a spontaneous breaking
of the residual axial symmetry, there are the domail wall excitations
interpolating between the above $N_c$ chirally asymmetric vacuums. 
The purpose of this section is to give a new qualitative description and
interpretation of the domain wall properties and, in particular, their 
ability to screen the quark charge. 

Let us recall in short the interpretation of the vacuum energy density 
behaviour in the pure gluodynamics which was proposed in \cite{I}.

The vacuum state at $\theta=0$ is supposed to be the condensate of pure
magnetic monopoles, i.e. the dyons with the magnetic and electric charges
$d_1^{\theta=0}=(1,\,0)$ (and of all $N_c-1$ types, as there are $N_c-1$ 
types of monopoles due to $SU(N_c)\ra U(1)^{N_c-1}$). 
 
As has been shown by E. Witten long ago \cite{w1}, 
as $\theta$ becomes nonzero the monopoles turn into the dyons with the
charges: $d_1^{\theta}=(1,\,\theta/2\pi)$. For this reason, the vacuum
energy density, ${\ov E}_{vac}(\theta)$, increases. 
This continues up to $\theta\ra \pi$ where
the above dyons look as: $d_1^{\theta=\pi}=(1,\,1/2)$. The above vacuum 
becomes unstable in the infinitesimal vicinity of $\theta=\pi$ because there 
is another state, the condensate of $d_2^{\theta=\pi}=(1,\,-1/2)$ - dyons,
degenerate in energy with the first one. Thus, there occurs rearragement of
the electrically charged degrees of freedom to recharge the $d_1$ - dyons 
into the $d_2$ - ones. For instance, a copious "production" of the charged
gluons, ${\bar g}=(0,\,-1)$ takes place, so that: $(\bar g)+d_1\ra d_2$.
This rechargement allows the system to have a lower energy at $\theta>\pi$.
Indeed, there are now only the $d_2^{\theta}=(1,\,-1+\theta/2\pi)$ - dyons 
in the condensate at $\theta>\pi$, their electric charge decreases with
increasing $\theta$ and the vacuum energy density decreases with it. As
$\theta \ra 2\pi$, the $d_2^{\theta}$ - dyons become the pure monopoles, and 
the vacuum state becomes exactly as it was at $\theta=0$. On the whole, the
vacuum energy density, ${\ov E}_{vac}(\theta)$, increases in some way at 
$0\leq \theta\leq \pi$; there is a cusp due to the above described 
rechargement at $\theta=\pi$, and it decreases then (in a symmetric way)
reaching its minimal value at $\theta=2\pi$.

Now, let us return to SYM and let us suppose that we have integrated out all,
but the $\rho$ and $\phi\,\, (\,\lambda\lambda\sim \rho \exp \{i\phi\}\,)$ 
fields (really, we expect the field $\rho$ is unimportant for a qualitative
picture discussed below and we will ignore it, supposing simply that it
takes its vacuum value ${\bar \rho}\sim\Lambda^3$).

As the field $N_c\phi$ in SYM is the exact analog of $\theta$ in YM, the
above described interpretation of the behaviour of ${\ov E}_{vac}(\theta)$ 
in YM can be transfered to SYM, with only some evident changes: a) ${\ov E}_
{vac}(\theta)\ra U(N_c\phi)$, and it is not the vacuum energy density 
now but rather the potential of the field $\phi$\,; 
b) if we start with the condensate of pure monopoles at $\phi=0$, the
rechargement $d_1^{\phi}=(1,\,N_c\phi/2\pi)\ra d_2^{\phi}=(1,\,-1+N_c
\phi/2\pi)$ and the cusp in $U(N_c\phi)$ will occur now at $\phi=\pi/N_c$, 
so that at $\phi=2\pi/N_c$ we will arrive at the next vacuum with the same 
pure monopole condensate but with the shifted phase of the gluino condensate.  

Let us consider now the domain wall exitation, $\phi_{dw}(z)$, interpolating 
along the "z" axis between, say, two nearest vacuums: $\phi(z\ra -\infty)\ra 
0$ and $\phi(z\ra\infty)\ra 2\pi/N_c.$ There is a crucial difference between 
this case and those just described above where the field $\phi$ was 
considered as being space-time independent, i.e. $\phi(z)=const$. The matter 
is that the system can not behave now in a way described
above (which allowed it to have a lowest energy at each given value of 
$\phi(z)=\phi=const$): i.e. to be the pure condensate of $d_1^{\phi}$ -
dyons at $0\leq \phi<\pi/N_c$, the pure condensate of $d_2^{\phi}$ - dyons at 
$\pi/N_c < \phi\leq 2\pi/N_c$, and to recharge suddenly at $\phi=\pi/N_c)$. 
The reason is that the fields corresponding to electrically charged degrees 
of freedom also become functions of "z" at $q=\int dz\, [ d\phi_{dw}(z)/dz ]
\neq 0$. So, they can not change abruptly now at some $z=z_o$ where $\phi_
{dw}(z)$ goes through $\pi/N_c$,
because their kinetic energy will become infinitely large in this case. Thus, 
the transition will be smeared necessarily.

The properties of the domain wall in the pure gluodynamics with $\theta=\pi$
were described in \cite{I}.
The properties of the domain wall under consideration here will be similar 
to those described in \cite{I}. The main difference is that $\theta$
was fixed at $\pi$ in \cite{I}, while $N_c\phi_{dw}(z)$ varies here smoothly 
between its limiting values, and the electric charges of dyons follow it.

So, at far left there will be a large coherent condensate of $d_1^{\phi}=
(1,\,N_c\phi/2\pi)$-dyons and a small (incoherent) density of $d_2^{\phi}=
(1,\,-1+N_c\phi/2\pi)$-dyons.
\footnote{
Other possible dyons play no role in the transition we consider, and we will
ignore them.
}
The $d_2^{\phi}$-dyons can not move freely in this region as they are on the 
confinement and appear as a rare and tightly connected pairs, ${\ov d}_2^
{\phi}d_2^{\phi}$, only. Therefore, their presence does not result in the 
screening of the corresponding charge. As we move to the right, the density
of $d_1^{\phi}$-dyons decreases while those of $d_2^{\phi}$ - increases.
These last move more and more freely, but are still on the confinement.
Finally, their density reaches a critical value so that a "percolation" takes
place, and the $d_2^{\phi}$-dyons form a 
continuous coherent network within which 
individual $d_2^{\phi}$-dyons can move freely to any distance. At the same 
time, there still sirvives a sufficiently large coherent condensate of 
$d_1^{\phi}$-dyons, which still can freely move individually within their own 
network. 

At the symmetrical place to the right of the domain wall centre the
"inverse percolation" takes place, so that the network of $d_1^{\phi}$-dyons 
decays into separate independently fluctuating droplets, whose density 
(and size) decreases with further increasing z. At large z we arrive at the 
vacuum state with a large coherent condensate of monopoles 
(former $(1,\,-1)$-dyons at large negative z). 

The above described system has some features in common with the mixed state
of the type-II superconductor in the external magnetic field. The 
crucial difference is that the magnetic flux tubes are sourceless inside the
superconductor, while in our case there is a finite density of freely
moving real charges (and anticharges) within each network.

Each time when there will coexist the condensates of two mutually non-local 
fields, they will try to keep each other on the confinement, and will 
resemble the above described case.

For instance, in SUSY SU(2)-YM with one matter flavour, there will be three
phases, depending on the value of $m,$ - the mass parameter of the (s)quark. 
At small $m<m_1=C_1\Lambda$, there will be the usual electric Higgs phase, 
with the magnetic charges being on the confinement, and with the monopoles 
appearing as independently fluctuating neutral droplets only. At $m=m_1$ the 
"percolation" of the monopole droplets takes plase, so that at $C_1<(m/
\Lambda)<C_2$ there will be "the double Higgs phase" with two coherent 
networks of monopoles and electric Higgs particles, with their averaged 
densities being constant over the space, and following only the value of $m$.
\footnote{
\,\,The kinetic terms of magnetically charged fields will have peculiarities 
at $m=m_{1}$, while the point $m=m_2$ will be special for the kinetic terms 
of electrically charged fields.
}
There will be screening rather than confinement (although the difference 
between these becomes somewhat elusory here) of any test charge in this 
interval of $m$. Finally, at $m=m_2=C_2\Lambda$ the "inverse percolation" of 
the electric Higgs condensate takes place, so that 
there will be only independently fluctuating neutral droplets of the electric
Higgs particles at $m>m_2$, and the usual confinement of the electric charge.

Now, let us return to our original theory and
consider what happens when a heavy quark is put inside the bulk of 
the domain wall. The crucial point is that there is a mixture of all four 
dyon and antidyon species (of all $N_c-1$ types): $d_1^{\phi}=(1,\,N_c\phi/2
\pi),\, {\bar d}_{1}^{\phi}=(-1,\,-N_c\phi/2\pi),\,d_2^{\phi}=(1,\,-1+N_c
\phi/2\pi)\,$ and ${\bar d}_2^{\phi}=(-1,\,1-N_c\phi/2\pi)$ in this 
"percolated region", with each dyon moving freely inside its coherent
network. So, this region has the properties of "the double Higgs phase", as
here both the $d_1^{\phi}$ and $d_2^{\phi}$-dyons are capable to screen 
corresponding charges. And because the charges of $d_1^{\phi}$ and $d_2^
{\phi}$-dyons are linearly independent, polarizing itself appropriately
this mixture of dyons will screen any test charge put inside, the heavy 
quark one in particular.

If the test quark is put at far left (right) of the wall, the string will 
originate from this point making its way toward a wall, and will disappear 
inside the bulk (i.e. the region of the double Higgs phase) of the wall. 
The above described explanation differs from both, those described by E. 
Witten in \cite{w2} and those proposed by I. Kogan, A. Kovner and M. Shifman 
in \cite{KKS}. 
\vspace{.5cm}
\begin{center} {\bf Acknowledgements} \end{center}

This work was supported in part by the Cariplo Foundation for Scientific
Research in collaboration with Landau Network-Centro Volta. I am deeply
indebted the whole staff, and espessially to A.Auguadro, G.Marchesini and
M.Martellini for a kind hospitality extended to me during my stay at Milano.

The work is supported in part by RFFR, \# 96-02-19299-a.

\end{document}